# Identifying Aspects for Web-Search Queries

**Fei Wu** WUFEI@GOOGLE.COM
**Jayant Madhavan** JAYANT@GOOGLE.COM
**Alon Halevy** HALEVY@GOOGLE.COM
*Google Inc, 1600 Amphitheatre Pkwy,*
*Mountain View, CA 94043 USA*

## Abstract

Many web-search queries serve as the beginning of an exploration of an unknown space of information, rather than looking for a specific web page. To answer such queries effectively, the search engine should attempt to organize the space of relevant information in a way that facilitates exploration.

We describe the ASPECTOR system that computes *aspects* for a given query. Each aspect is a set of search queries that together represent a distinct information need relevant to the original search query. To serve as an effective means to explore the space, ASPECTOR computes aspects that are orthogonal to each other and to have high combined coverage.

ASPECTOR combines two sources of information to compute aspects. We discover candidate aspects by analyzing query logs, and cluster them to eliminate redundancies. We then use a mass-collaboration knowledge base (e.g., Wikipedia) to compute candidate aspects for queries that occur less frequently and to group together aspects that are likely to be "semantically" related. We present a user study that indicates that the aspects we compute are rated favorably against three competing alternatives – related searches proposed by Google, cluster labels assigned by the Clusty search engine, and navigational searches proposed by Bing.

## 1. Introduction

Web-search engines today predominantly answer queries with a simple ranked list of results. While this method has been successful, it relies on the assumption that the user's information need can be satisfied with a *single* page on the Web. However, several studies (Broder, 2002; Rose & Levinson, 2004) have alluded to the fact that many user queries are merely the beginning of an exploration of an unknown space of information. Such queries are likely to be better served if users were provided with a summary of the relevant information space and the means for conveniently exploring it. To paraphrase, rather than finding the *needle* in the haystack, some queries would benefit from *summarizing* the haystack (Rajaraman, 2008). Several commercial attempts have recently been made to provide better answers to such queries, including systmes like CARROT2, CLUSTY, KOSMIX, and YAHOO!GLUE.

This paper describes the ASPECTOR system that addresses the following problem: given an exploratory search query $q$, compute a set of *aspects* that enables convenient exploration of all the Web content that is relevant to $q$. We define each aspect as a set of search queries that together represent a distinct information need relevant to the original search query, similar to Wang's definition of Latent Query Aspect (Wang, Chakrabarti, & Punera, 2009). For example, consider the queries in Table 1 and their potential aspects. Each aspect covers a different kind of information and together they span a large amount of the relevant information that search engine users might be interested in. Two simple ways a search engine can utilize aspects is to offer them as related searches or to categorize search results by the aspects they are most relevant to. Aspects can also form the basis for various mashup-like interfaces, e.g., the aspect pictures can trigger the inclusion of images, while weather can trigger a weather report gadget. Computing aspects for queries can be seen as a first step towards mining a knowledge base, called the *database of user intentions* (Battelle, 2005). This knowledge base is a timely and culture-sensitive expression of peoples' interests. Inferring such a knowledge base for entities can serve as a basis for an effective presentation of information, and can therefore have significant ramifications on search, advertising and information dissemination.

| **vietnam travel** | **kobe bryant** |
|---|---|
| travel guides | statistics |
| packages / agencies | pictures / photos |
| visa | videos / youtube |
| blogs / forums | shoes |
| travel advisories | injury reports |
| weather | girlfriend |
| cities (Hanoi / Saigon /...) | trade rumors |

Table 1: Potential aspects for the queries vietnam travel and kobe bryant.

Aspector computes aspects for a query $q$ using a search-engine query log, augmented with information from a knowledge base created by mass-collaboration (Wikipedia). Given a query $q$, related queries are extracted from the query log as candidate aspects. While the logs are an excellent mirror of users' interests, they can also result in noisy and redundant aspects, e.g., top related queries for vietnam travel include vietnam visa and vietnam travel visa. Furthermore, query logs are of limited utility for generating aspects for less popular queries, e.g., there are much fewer related queries for laos travel than vietnam travel. We describe the following algorithmic innovations that address these challenges. First, we show how redundant candidate aspects can be removed using search results. Second, we apply *class-based label propagation* in a bipartite graph to compute high-quality aspects even for a long tail of less popular queries. Finally, we show that knowledge bases can be used to group candidate aspects into categories that represent a single information need. We believe that our solution demonstrates an interesting interplay between query logs and knowledge bases that has not as yet been investigated in the research literature.

We describe a detailed experimental evaluation of Aspector. We compare the aspects generated by Aspector against three possible competing approaches – related searches proposed by Google.com, cluster labels proposed by Clusty.com, and navigational searches proposed by Bing.com. Related searches and navigational searches are typically also generated by the analysis of query logs. Cluster labels are generated by grouping the search results of the original query and extracting labels from the documents within each cluster. We show that our aspects are more diverse than all other three systems. We also show that our aspects span a larger space of information – not only do they expose more results than the original query, but the additional results are considered highly relevant by users. Our user study finds that the results of Aspector are preferred over related searches, cluster labels and navigational searches as a means for further exploration.

Section 2 defines the problem of computing aspects, Section 3 considers potential alternative approaches. Section 4 describes the generation of candidate aspects, and Section 5 describes how Aspector selects aspects from the candidates. Section 6 describes our experimental evaluation,and Section 7 describes related work. Section 8 concludes.

## 2. Problem Definition

We begin by defining the scope of the problem we address.

**Queries:** We assume that queries are a sequence of keywords, as is typical for search-engine interfaces. Our techniques are not meant to apply to arbitrary queries. We focus on exploratory queries, and specifically, we assume that they either are *entity names* (e.g., the country Vietnam) or have an *entity and a property name* (e.g., Vietnam travel). Thus, we are interested in computing aspects for entities in general or in the context of a particular property.

In this paper we do not handle the problem of segmenting entity and property names in queries (previous work, such as Bergsma & Wang, 2007; Tan & Peng, 2008, have addressed the problem). The question of identifying exploratory queries in a query stream is also beyond the scope of this paper.

**Aspects:** An aspect of a query $q$ is meant to describe a particular sense of $q$ corresponding to an information need. Specifically, each aspect is represented by a collection of search queries related to $q$. Given $q$, we compute a set of aspects $a_1, \ldots, a_n$, along with scores $p(a_i|q)$ that can then be used to rank them.

Since aspects are collections of search queries, we compare aspects based on the search results retrieved by the queries that constitute them. Aspects are meant to capture diverse dimensions along which we can organize the exploration of the entire space of information relevant to query. Hence, the set of aspects computed for each query should have the following properties:

**Orthogonality:** given two aspects, $a_1$ and $a_2$, the search results for $a_1$ and $a_2$ should be very different from each other.

**Coverage:** the search results provided for the aspects should offer a good overview of the relevant space of information.

Thus, two *sets of aspects* computed for the same query can be compared based on the pairwise orthogonality of their constituent aspects and the combined coverage of all their aspects. The evaluation of aspects is inherently subjective, as is typical in the area of web search. Hence, we present user studies where the aspects computed by different approaches are qualitatively rated by a large number of independent users. We note that while we compare different approaches to computing aspects, we do not focus on the different ways they can be presented to users.

## 3. Alternative Approaches

Before we describe how Aspector generates aspects, we briefly mention two strawman approaches to the problem and explain why they are insufficient for our needs.

| Class | Wikipedia | | Query Log | |
|---|---|---|---|---|
| NBA Player | birth date | position | injury | pictures |
| | birth place | college | nba | wallpaper |
| | nationality | height(ft) | bio | salary |
| | draft year | draft | shoes | girlfriend |
| | career start | height(in) | stats | biography |
| University | name | city | library | basketball |
| | established | website | football | athletics |
| | country | type | alumni | admissions |
| | campus | state | tuition | baseball |
| | undergrad | motto | jobs | bookstore |

Table 2: Two classes that have attributes in Wikipedia that are very different from class-level aspects computed from the query log.

### 3.1 Community-Created Knowledge Bases

Knowledge bases, especially those created by a large community of contributors, are a rich source of information about popular entities. They cover a wide spectrum of user interests and can potentially be used to organize information relevant to search queries. In particular, the properties in Infoboxes of Wikipedia articles can potentially be used as candidate aspects. Wikipedia contains more than 3,500 classes with over 1 million entities and each class on average has about 10 attributes. The Wikipedia column in Table 2 shows the attributes for two example classes. Freebase is another community-created KB with over 1,500 classes.

Binary relationships recorded in a knowledge base fall significantly short of providing a good set of aspects for a query. For example, consider the properties associated with Cambodia in the Wikipedia Infobox – capital, flag, population, GDP, etc. None of these words appear in the top-10 frequent queries that contain the word cambodia. In addition, a knowledge base is limited to describing well defined entities. For example, Cambodia is an entity in a knowledge base, but Cambodia Travel is not. However, queries on the Web cover much more than well defined entities.

The underlying reason that knowledge bases fall short is that their constructors choose attributes based on traditional design principles, but good aspects do not follow from these principles. For example, it turns out that cambodia travel is a good aspect for vietnam travel, because many people consider a side trip to Cambodia when visiting Vietnam. However, when designing a knowledge base, Cambodia would never be an attribute of Vietnam. Instead, the knowledge base would assert that Vietnam and Cambodia are neighbors, and include a rule that states that if X is next to Y, and then X travel may be an aspect of Y travel. Unfortunately, coming up with such rules and specifying their precise preconditions is a formidable task and highly dependent on the instances it applies to. For example, pakistan travel is not an aspect of india travel, even though the two countries are neighbors.

### 3.2 Web Documents

Another approach to finding aspects is to cluster the documents on the Web that are relevant to a query $q$, and assign or extract labels for each cluster (Blei, Ng, & Jordan, 2003; Zeng, He, Chen, Ma, & Ma, 2004). As we show in our experiments, the main disadvantage is that the coverage of the resulting aspects may be low because this approach only considers documents that were returned in response to the original query. In practice, users conduct data exploration in *sessions* of queries, and the other queries in those sessions can also lead to interesting aspects which might not be found among the results of the original query. Furthermore, it can be very challenging to generate succinct names for aspects from each cluster.

## 4. Generating Candidate Aspects

Aspector generates candidate aspects from query logs. Query logs are very reflective of a broad range of user interests, but they are less effective in generating aspects for infrequent queries. We first describe how we generate *instance-level* aspects, and then how we augment them with *class-based aspect propagation* using a knowledge base.

### 4.1 Instance-Level Candidate Aspects

Given query $q$, we start by considering each of its query refinements and super-strings as a *candidate aspect*.

#### 4.1.1 Query Refinements

A query $q_j$ is a refinement of $q$, if a user poses $q_j$ after $q$ while performing a single search task. Query logs can be mined to identify popular refinements for individual queries. Search engines typically use popular refinements as a basis for proposing related searches.

We process refinements as follows: first, the query log is segmented into sessions representing sequences of queries issued by a user for a single search task. Suppose $f_s(q, q_j)$ is the number of sessions in which the query $q_j$ occurs after $q$, we then estimate the refinement score $p_r$ for each $q_j$ by normalizing $f_s(q, q_j)$ over all possible refinements, i.e.,

$$p_r(q_j|q) = \frac{f_s(q, q_j)}{\sum_i f_s(q, q_i)}$$

Observe that proposing related searches based on query refinements is, in principle, only optimized towards the goal of helping users find a single page containing a specific answer (rather than helping the user explore the space). For example, the top 10 refinements for the query on the NBA player yao ming includes 6 other NBA players such as kobe bryant and michael jordan. Though related, these refinements are not necessarily the best aspects for the query.

#### 4.1.2 Query Super-Strings

The query $q_j$ is a super-string of $q$ if it includes $q$ as a sub-string. For example, vietnam travel package is a super-string of vietnam travel. Unlike a refinement, a super-string $q_j$ need not belong to the same session as $q$. In fact, for a random selection of 10 popular queries, we found that on average there is only an overlap of 1.7 between the top 10 refinements and the top 10 super-strings. In a sense, super-strings are explicitly related queries while refinements are more implicitly related.

Super-strings can be assigned scores similar to $p_r$ above, by mimicking each super-string as a pseudo-refinement, i.e., we assume an imaginary session in which $q$ preceded super-string $q_j$. Suppose $f(q_j)$ was the number of occurrences of $q_j$ in the query logs, we estimate the super-string score $p_{ss}(q_j|q)$ as below [1]:

$$p_{ss}(q_j|q) = \frac{f(q_j)}{f(q) + \sum_i f(q_i)}$$

Aspector considers all the refinements and super-strings of $q$ as candidate aspects and assigns them a single instance-level aspect score. For each candidate aspect $q_j$, we assign the score $p_{inst}$ as follows:

$$p_{inst}(q_j|q) = max(p_r(q_j|q), p_{ss}(q_j|q))$$

For given $q$, we normalize all $p_{inst}(q_j|q)$s to add up to 1.

### 4.2 Class-Based Aspect Propagation

Query-log analysis is ineffective in generating instance-level candidate aspects for less frequent queries. For example, we generate good candidate aspects for vietnam travel, but not for laos travel. However, we can recommend aspects that are common to travel to many countries for Laos. We use a variation of the label-propagation algorithm named *Adsorption* (Baluja, Seth, Sivakumar, Jing, Yagnik, Kumar, Ravichandran, & Aly, 2008).

We first apply query segmentation to extract the entity $e$ (laos in our example) and the property $p$ (travel) from the query $q$. Next, we use a knowledge base (e.g., Wikipedia Infobox) to identify the class, or classes, $C$ of $e$ (e.g., country and south-east asian country for laos). Then we construct a directed bipartite graph $G = (V, E, \omega)$ as shown in Figure 1. The nodes on the left are instance-level query nodes such as "laos travel", and on the right are class-level nodes like "country travel". $E$ denotes the set of edges, and $\omega : E \to R$ denotes the nonnegative weight function. We set the weights of edges from instance nodes to class nodes as 1, and the weights of edges from class nodes to instance nodes as $K$, a design parameter controlling the relative-importance of the two factors. Our goal is to compute $\overrightarrow{p(q_j|q)}$, which is the aspect distribution on node $q$.

Following the work of Baluja et al. (2008), each node's aspect distribution is iteratively updated as the linear combination of its neighbors, until we converge (this algorithm is shown to be equivalent to performing a random walk in the graph). Since we use Wikipedia Infobox as our knowledge base, where each instance belongs to a single class, two iterations are guaranteed to achieve convergence. The first iteration computes the class-level aspects as follows:

$$p_{class}(q_j|q) = \frac{1}{|C|} \sum_{q \in C} p_{inst}(q_j|q)$$

1. We use a conservative lower bound estimate. The corresponding upper bound $p_{ss}(q_j|q) = \frac{f(q_j)}{f(q)}$ can exceed 1.

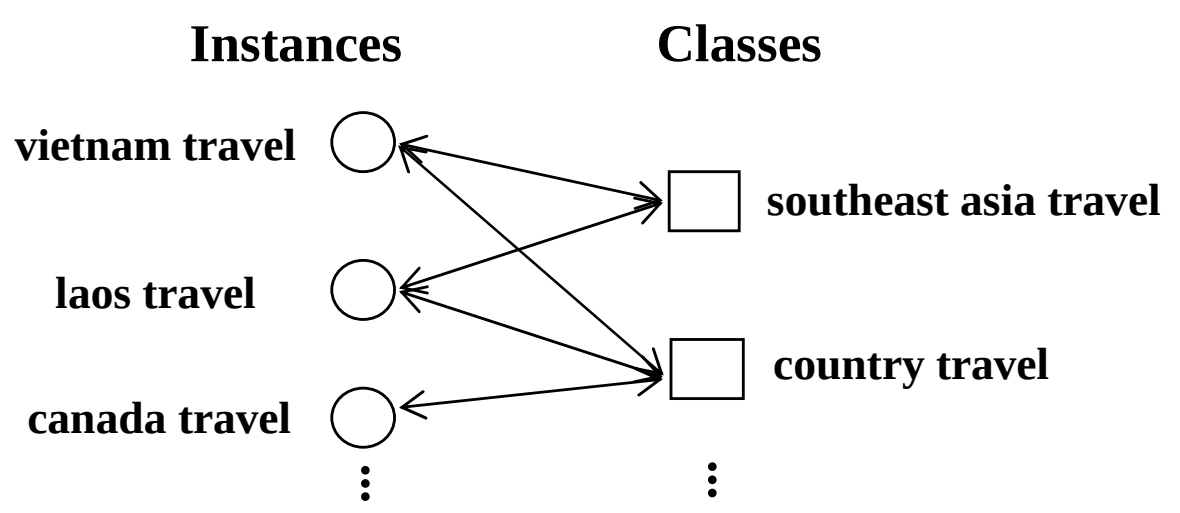

Figure 1: The bipartite graph for class-based aspect propagation.

The second iteration smoothes the aspect distribution on instance node $q$ with $p_{class}(q_j|q)$ as follows,

$$p(q_j|q) = \frac{p_{inst}(q_j|q) + K \times p_{class}(q_j|q)}{1+K} \quad (1)$$

In Section 6 we tested the effect of $K$ on the performance of Aspector. In our experiments we found that the following variation on computing class-level aspects leads to slightly better results:

$$p_{class}(q_j|q) = \frac{1}{|C|}\sum_{q\in C} I(p_{inst}(q_j|q) > 0))$$

where, $I(p_{inst}(q_j|q) > 0)) = 1$ when $p_{inst}(q_j|q) > 0$, and 0 otherwise.

Table 2 shows examples of top class-level aspects derived for two classes and compares them with their corresponding top attributes in the Wikipedia infobox. We can see that the two sets of aspects have little overlap, which illustrates again that community created schemata fall significantly short of providing a good set of aspects for search queries.

## 5. Selecting Aspects

This section describes how Aspector prunes the set of candidate aspects, groups them, and eventually ranks them in an ordered list from which a subset can be selected.

### 5.1 Eliminating Duplicate Aspects

It is often the case that the generated aspect list contains very similar candidates that may be considered redundant. For example, top candidate aspects for the query vietnam travel include vietnam travel package, vietnam travel packages and vietnam travel deal, each of which represent either identical or very similar user intents. In particular, note that the set of web documents returned for each of the aspects from any search engine are likely to be very similar.

To remove redundant aspects, we compute a similarity matrix, $\{sim(a_i, a_j)\}$, between every pair of candidate aspects and then cluster them based on their similarity.

#### 5.1.1 Computing Aspect Similarity

Since most aspects only contain few words, estimating similarity based on a simple comparison of words is unlikely to be accurate. Therefore, we enrich the representation of each aspect by considering the top $m^2$ search results returned when posing the aspect as a search query. This is consistent with our goal of enabling orthogonal exploration – aspects with similar top results are unlikely to be orthogonal.

Let $D_i$ be the top web pages retrieved for the aspect $a_i$. We estimate the similarity of $a_i$ and $a_j$ to be the similarity of the corresponding sets $D_i$ and $D_j$. To compute $sim(D_i, D_j)$, we first compute the similarity $dsim$ for any given pair of web pages $\{d_i \in D_i,\ d_j \in D_j\}$. For this we use the standard *cosine distance* between the TF/IDF word-vectors for the two documents. For computational efficiency, we only consider the head and snippet for each web page instead of their entire text contents[3].

While $sim(D_i, D_j)$ can potentially be estimated by averaging similarities $dsim(d_i, d_j)$ for all pairs of web pages, on our experiment dataset, we found it better instead to compute the average of the highest similarity for each web page. For each $d_i \in D_i$, we assign the score: $sim(d_i, D_j) = max_k dsim(d_i, d_k)$. Likewise, we assign $sim(D_i, d_j) = max_k dsim(d_k, d_j)$. The final aspect similarity is computed as:

$$sim(a_i, a_j) = sim(D_i, D_j) = \frac{\sum_i sim(d_i, D_j)}{2|D_i|} + \frac{\sum_j sim(d_j, D_i)}{2|D_j|}$$

We could alternatively treat each $D_i$ as one single document by concatenating all $\{d_i \in D_i\}$ and estimate $sim(q_i, q_j)$ to be the corresponding $dsim(D_i, D_j)$. While computationally more efficient, the quality of aspects was poorer.

#### 5.1.2 Clustering Aspects

In principle, we can apply any clustering algorithm, such as K-means or spectral clustering, to the resulting aspect similarity matrix. However, these algorithms often require pre-setting the number of desired clusters, which is difficult in our context. In addition, the number of clusters also varies significantly from one query to another. Note that the appropriate number of clusters is *not* necessarily the number of resulting aspects that we will show the user.

We instead apply a graph-partition algorithm for clustering. The algorithm proceeds by creating a graph where the nodes are aspects, $a_i$, and there is an edge connecting the nodes $a_i$ and $a_j$ if $sim(a_i, a_j) > \sigma$, where $\sigma$ is a pre-defined threshold. Each of the connected sub graphs is treated as a cluster. We choose the label of the cluster to be the aspect $a_k$ with the highest $p(a_k|q)$ in the cluster (Formula 1).

The only design parameter $\sigma$ is easier to set and pretty stable for different queries, as shown in our experiments. We note that similar algorithms such as star-clustering (Aslam, Pelekov, & Rus, 2004) can also be used.

### 5.2 Grouping Aspects by Vertical Category

In many cases, even after eliminating redundant aspects, we find that we are left with aspects that are seemingly different, but can be semantically grouped into a single category. For example, for the query `vietnam travel`, some of the top non-redundant candidate aspects

2. We use $m = 8$ in our experiments which performs well. Larger $m$ might achieve slightly better performance at the cost of heavier computation.
3. We also tired using the whole document for each web page, which has only slightly better performance.

as ho chi minh city, hanoi and da nang. While these are different cities, in principle they are likely to represent a single information need – that of finding more information about cities in Vietnam. Further, given a budget of a fixed number of aspects that can be presented to a user, it might not make sense to overwhelm them with a list of aspects all denoting cities in Vietnam. Instead, a single aspect named Cities can be presented.

Here is where the community-created knowledge bases can be leveraged – Aspector tries to identify sets of related aspects by consulting the Wikipedia Infobox system[4]. If it finds that multiple aspects contain different entities that belong to a class in Wikipedia, it creates an aggregate aspect (with the label of the class) and groups them together.

We encounter two challenges while looking up Wikipedia for the classes of entities. First, the same entity can appear as different synonymous tokens. For example, nyu is the common acronym for new york university. Currently we use the redirect pages on Wikipedia to infer synonyms. Redirect pages in Wikipedia point synonym terms to the same principal article. As a result, the aspect nyu for the query yale university is grouped with harvard university and oxford university[5]. Second, the same token can refer to multiple entities that belong to different classes and it can lead to bad grouping decisions. For example, HIStory and FOOD are the names of music albums in Wikipedia, but history and food are also aspects for the query vietnam. A simple lookup of the tokens in Wikipedia might lead to erroneously grouping them into a single album group. Aspector uses the disambiguation pages in Wikipedia to identify tokens that are likely to have multiple senses. The Infobox class is only retrieved for entities that do not have disambiguation pages. This conservative method can be further improved via collaborative classification (Meesookho, Narayanan, & Raghavendra, 2002). For example, earth, moon and venus are all aspects for mars. Since all of them are ambiguous based on Wikipedia, our current Aspector would treat them as individual aspects. However, it is possible to group them together as a single planet aspect, given all three candidates have planet as one possible type.

### 5.3 Selecting Aspects

The final step in Aspector is selecting the aspects. We note that absolute ranking of aspects is not so important in our context, because we expect that search results from aspects will be spread out on the screen rather than being presented as a single list. However, we still need to select the top-k aspects to present. Our selection of top-k aspects is based on our original goals of increasing coverage and guaranteeing orthogonality.

Aspector uses the score of an aspect, $p(a_i|q)$, as a measure of coverage. To achieve a balance between coverage and orthogonality, Aspector uses a greedy algorithm that selects aspects in the ratio of their score $p(a_i|q)$ and the similarity of the aspects to already selected aspects. The algorithm below produces a ranked list of aspects, $G$.

4. Other ontologies like Freebase and Yago can also be used.
5. This trick is used when constructing the bipartite graph in Section 4.2 as well.

**Input**: Set $S = \{a_i\}$ `// Label aspects of clusters after de-duplication.`
**Output**: Set $G$ `// Ranked list of aspects.`

Initialization: $G = \phi$;

$a_0 = argmax_{a_i \in S} p(a_i|q)$;
move $a_0$ from $S$ to $G$;

**while** $(S \neq \phi)$ **do**
 **for** $a_i \in S$ **do**
  set $sim(a_i, G) = max_{a_j \in G} sim(a_i, a_j)$;
 $a_{next} = argmax_{a_i \in S} \frac{p(a_i|q)}{Sim(a_i,G)}$;
 move $a_{next}$ from $S$ to $G$;

**Algorithm 1**: Aspector selects top-k aspects by balancing coverage and orthogonality.

Observe we set the similarity $sim(a_i, G)$ to be the maximum similarity of $a_i$ to the aspects already in $G$. On termination, Aspector returns the top $n$ aspects in ranked order (in our experiments we used $n = 8$). Our experiments indicate that balancing coverage and orthoganality leads to better selection of aspects than simply using coverage.

## 6. Experiments

In this section we evaluate our system Aspector and in particular, answer the following questions.

**Quality of aspects:** We compare the results of Aspector against three potential competing systems – related searches proposed by Google (henceforth Grs), cluster labels assigned by the Clusty search engine (Ccl), and navigational searches proposed by Bing (Bns). To better support exploration of different parts of the space of relevant information, the aspects of a query have to be orthogonal to each other. Aspects should also increase the coverage, i.e., reveal information that is not already available through the original query, but is still very relevant to it. Using a combination of search result analysis and a user study, we show that our aspects are less similar to each other (and hence more orthogonal) (Section 6.3), that aspects are able to increase coverage (Section 6.4), and that aspects are overall rated more favorably than Grs, Ccl, and Bns (Section 6.5).

**Contributions of the different components:** Aspector generates instance-level aspects and performs class-based aspect propagation, eliminates duplicates, and groups the remaining ones using a knowledge base. We show that instance-level and class-level aspects tend to be very different, and that the best results are obtained by judiciously combining them (Section 6.6). We also show that our clustering algorithm is able to stably eliminate duplicate aspects crossing different domains, and the grouping of aspects has a positive impact on the quality of aspects (Section 6.7).

### 6.1 Experimental Setting

To compute candidate aspects from query logs, we used three months worth of anonymized search logs from Google.com. We used a snapshot of the English version (2008.07.24) of the Wikipedia Infobox to serve as our knowledge base. Unless otherwise mentioned, we used $K = 0.1$ for class-based aspect propagation (Equation 1). We now describe our test suite and our user study.

**Test Queries:** We focus on queries that are entity names or have an entity name and a property name. We construct a test suite that contains 6 sets of queries: five with entity names from the Wikipedia classes Country, NBA player, Company, Mountain, and University, and one with entity-property queries of the form Country travel. To construct a mix of popular and rare queries, in each of the six sets we select 5 queries that occur frequently in the query stream, 5 that are relatively uncommon, and 5 are chosen randomly for the class (as long as they appear in the query logs). Thus, in total we have 90 test queries. For each experiment we used a random subset of these test queries.

**User Study:** As part of our experimental analysis, we performed user studies using the Amazon Mechanical Turk (Amt) system. On Amt, requesters (like us) post tasks and pay for anonymous registered workers to respond to them. Tasks are structured as a sequence of questions that workers are expected to respond as per the instructions provided by the requester. For example, to compare two algorithms that compute aspects, we can design a sequence of tasks such that in each a query and two lists of aspects (computed by each algorithm) are shown. The worker has to rate whether one list is better than the other or they are very similar. Amt ensures that each worker can only respond to a task once. Since, the workers in the user study are completely unknown to the requester, there is less of chance of bias. Amt has been shown to be an effective and efficient way to collect data for various research purposes (Snow, O'Connor, Jurafsky, & Ng, 2008; Su, Pavlov, Chow, & Baker, 2007). In our experiments, we used the default qualification requirement for workers that requires each worker to have a HIT approval rate (%) greater than or equal to 95.

### 6.2 Points of Comparison

Grs can be considered as a representative of current approaches that are based on mining refinements and super-strings from query logs. It is likely that Grs only performs an instance-level analysis and it does not attempt to identify distinct user information needs.

Ccl clusters result pages and assigns human-understandable labels to each cluster. Most notably, the clusters are determined purely from results of the original query, and there is no attempt to enable exploration of results that were not retrieved by the query. Further, it is likely that cluster labels are extracted by an analysis of the contents of the result pages (web documents). We note that while the clustering is hierarchical, for our experiments we only considered the top-level labels.

Bns provides navigation searches (right above the "Related searches" in the result pages) to help users better explore the information space. Bing's goal is closest to ours in spirit, but their technique only applies in a narrow set of domains. We note that Bns sometimes provides generic aspects (e.g., videos, images), but we do not consider those.

We note that neither Grs nor Ccl were designed with the explicit goal of computing aspects that help explore the information space relevant to a query. However, they can be viewed as close alternatives in terms of the results they may produce, and therefore offer two points of comparison.

Table 3 shows the aspects, related searches, cluster labels, and navigational searches obtained by the four systems on some example queries. In the rest of this section, we will first show that aspects from Aspector are on average more orthogonal, increase coverage, and are rated better overall than Grs, Ccl and Bns.

| Query | Grs | Ccl | Bns | Aspector |
|---|---|---|---|---|
| Mount Shasta | volcano<br>national park<br>climbing<br>vortex<br>camping<br>hotels<br>attractions<br>lodging | photos<br>hotels<br>real estate<br>weed<br>wilderness, california<br>climbing<br>weather, forecast<br>ski | image<br>weather<br>real estate<br>hotels<br>lodging<br>rentals<br>reference/wikipedia | resort<br>weather<br>high school<br>real estate<br>hiking<br>pictures (photos)<br>map<br>ski area |
| Yale University | harvard university<br>athletics<br>press<br>brown university<br>stanford university<br>columbia university<br>cornell university<br>duke university | school<br>department<br>library<br>images<br>publications<br>admissions<br>laboratory<br>alumni | admissions<br>jobs<br>bookstore<br>alumni<br>library<br>reference/wikipedia<br>images | press<br>art gallery<br>athletics<br>harvard (oxford, stanford,...)<br>jobs<br>bookstore<br>admissions<br>tuition |

Table 3: Sample output from Grs, Ccl, Bns, and Aspector.

### 6.3 Orthogonality of Aspects

To establish the orthogonality of aspects, we measure the *inter-aspect* similarity – the less similar the aspects are, the more orthogonal they are. We first describe how we compute inter-aspect similarity, and then report its values over the query set for Aspector, Grs, Ccl, and Bns.

In Section 5, we used TF/IDF-based word vectors to estimate aspect similarity. Using the same measure to establish orthogonality will bias the evaluation in favor of Aspector. Hence, we use an alternate measure for aspect similarity that employs a *topic model* (Blei et al., 2003). Briefly, topic models are built by learning a probability distribution between words in documents and the topics that might underlie a document. Given a text fragment, a topic model can be used to predict the probability distribution of the topics relevant to the fragment. For example, the text on the company page for Google Inc., might result in the topic distribution $\langle\langle\text{search engine}, 0.15\rangle, \langle\text{online business}, 0.08\rangle, \ldots\rangle$. We use a topic model developed internally at Google (henceforth TMG). Given two text fragments $t_1$ and $t_2$, we can compute their topic similarity $tsim(t_1, t_2)$ as the *cosine distance* between their topic distribution vectors $\vec{T_1}$ and $\vec{T_2}$.

Since aspects contain only a few words, we extend augmenting each aspect with its corresponding top search results (as in Section 5). Given aspects $a_1$ and $a_2$, let $D_1$ and $D_2$ be their respective top $m$ web search results. We compare $D_1$ and $D_2$ using TMG to estimate aspect similarity. Specifically, we compute the average inter-document similarity.

$$sim(a_1, a_2) = \frac{1}{k^2} \sum_{d_i \in D_1, d_j \in D_2} tsim(d_i, d_j) \qquad (2)$$

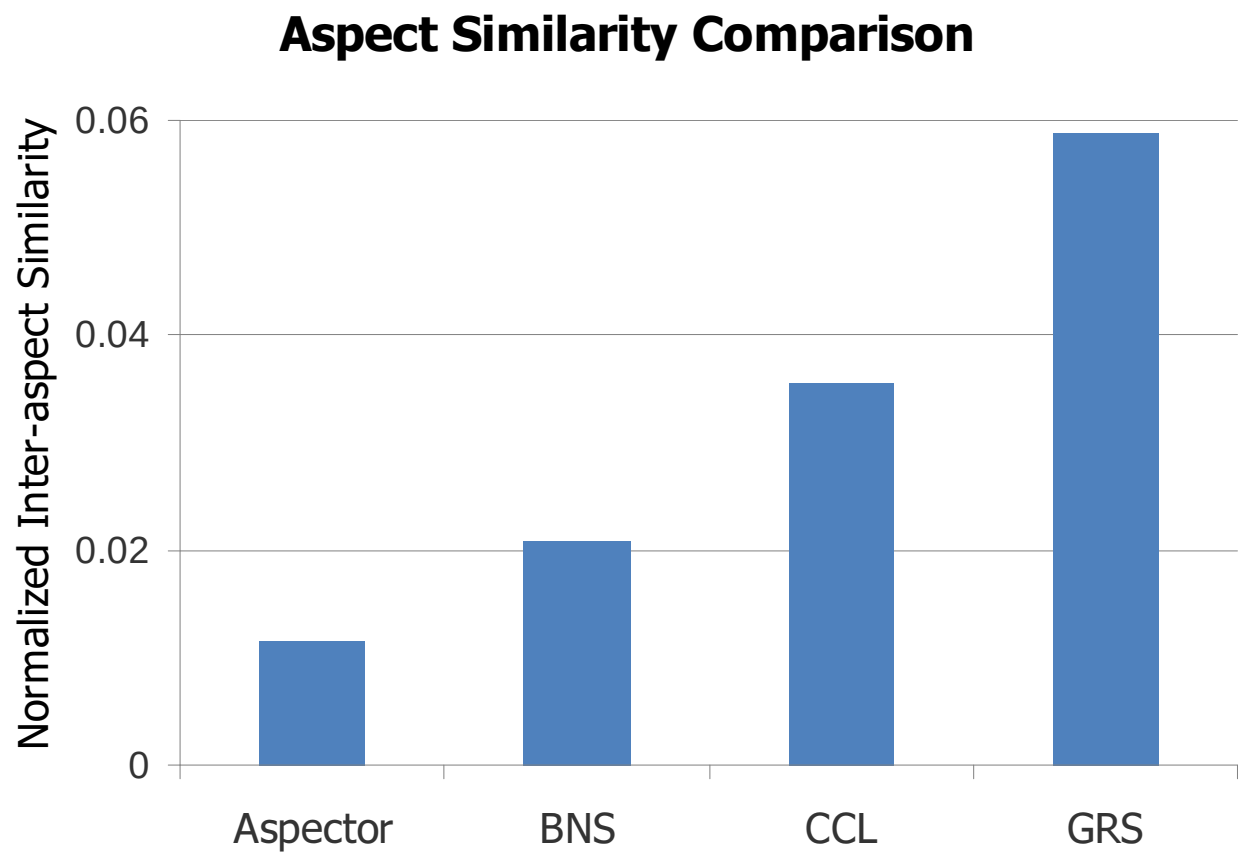

Figure 2: The results of Aspector are more orthogonal than those of Grs, Ccl, and Bns.

Given $A$, a set of $n$ aspects, we determine its *inter-aspect* similarity ($asim$) as the average pair-wise aspect similarity.

$$asim(A) = \frac{2}{n(n-1)} \sum_{a_i, a_j \in A} sim(a_i, a_j)$$

In order to make sense of the magnitude of $asim$, we normalize it using the average *intra-aspect* similarity $isim(A)$ obtained by comparing each aspect against itself.

$$isim(A) = \frac{1}{|A|} \sum_{a_i \in A} sim(a_i, a_i)$$

Note, $sim(a_i, a_i)$ is ususally not equal to 1 based on equation 2. The result is the *normalized inter-aspect* similarity $nsim$.

$$nsim(A) = \frac{asim(A)}{isim(A)}$$

Thus, if all the aspects in $A$ are identical, $nsim(A) = 1$, and if they are entirely orthogonal $nsim(A) = 0$.

For each query, we retrieved the same number of aspects (at most 8) from each system, and Figure 2 shows the average normalized inter-aspect similarity for the results output by each system.

As can be clearly seen, Aspector has the least normalized inter-aspect similarity and hence the most orthogonal aspects. The improvement over Bns (45%) is most likely due to the grouping of related aspects by vertical category. The improvement over Grs (90%) is most likely due to inclusion of class-based aspects and the grouping of related aspects by vertical category. The improvement over Ccl (60%) is likely because their space of labels is restricted to only the results returned by the original query.

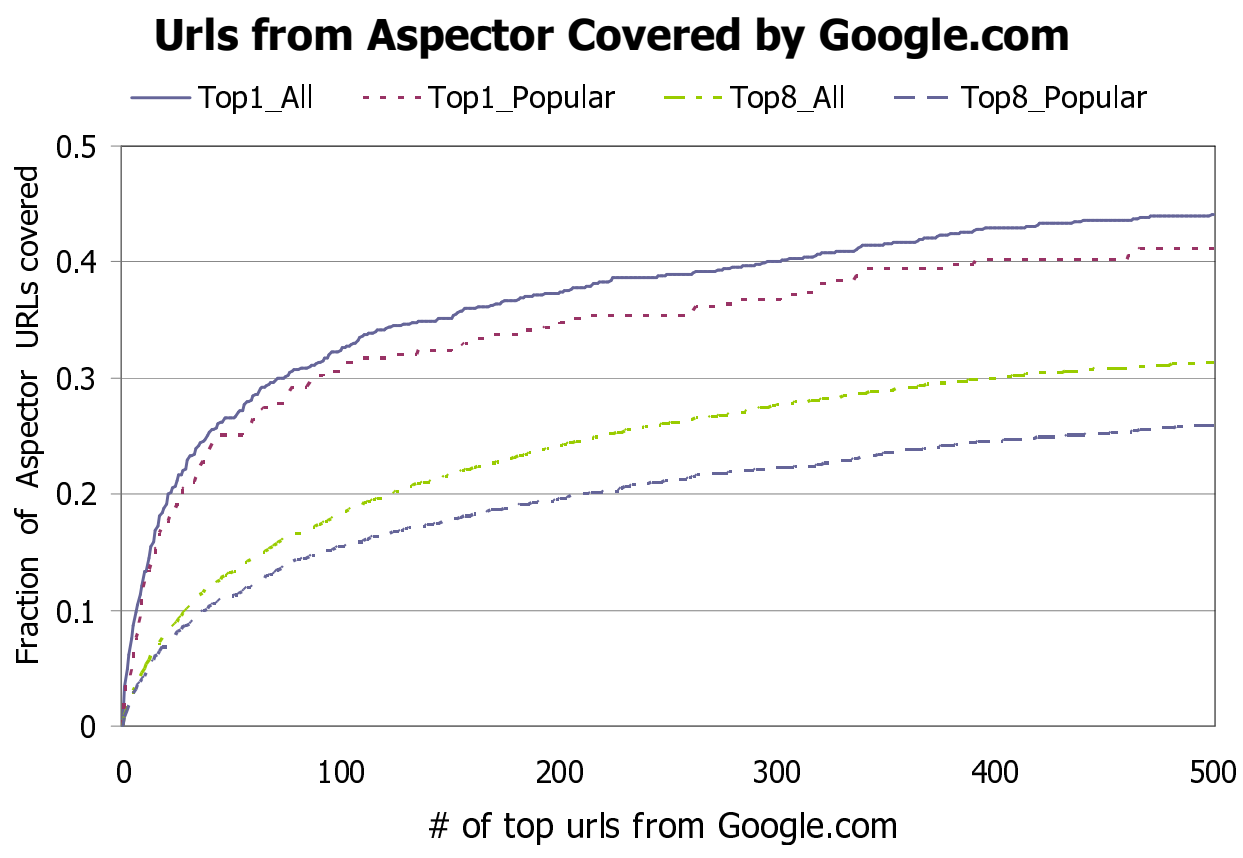

Figure 3: Fraction of top web pages retrieved by aspects that are also in the top 500 pages retrieved by the original search query.

### 6.4 Increase in Coverage

To validate the increase in coverage we are interested in answering two questions: (1) do the aspects enable users to reach more information than the original query? (2) is the additional information relevant to the user query?

#### 6.4.1 More Information

To show that aspects can reach more information, we compare the web pages retrieved using the aspects computed by Aspector against those retrieved by the original search query. Given, a query $q$ and its computed aspects $A$, let $D_N$ be the set of top $N$ web pages retrieved by Google for the query $q$. Let $D_k^i$ be the collection of top $k$ web pages retrieved by Google for an aspect $a_i$ in $A$, and let $D_k^a$ be the union of all the $D_k^i$s. We measure the fractional overlap between $D_N$ and $D_k^a$, i.e., $\frac{|D_k^a \cap D_N|}{|D_k^a|}$.

Figure 3 shows the average fractional overlap between $D_k^a$ and $D_N$ for $k = 1$ and $k = 8$ against different values of $N$ (x-axis). The results are averaged over two sets of queries: (1) all 90 queries, and (2) a subset of 30 popular queries, with 10 aspects being computed for each query. As the results clearly indicate, even when considering the top 500 search engine results, for $k = 1$, only about 45% of the web pages in $D_1^a$ are retrieved. In other words, 55% web pages retrieved using the aspects are not even in the top 500 (note that $|D_1^a|$ is only 10). The overlap is an even lower 33% when considering $k = 8$. This shows that aspects are clearly able to retrieve more new information.

In order to isolate the potential effects due to rare queries for which search engines do not typically propose related searches, we separately consider the subset of popular queries. Interestingly, here we find that the overlaps are even smaller and hence aspects are able to retrieve even more information. This is likely because there is potentially more diverse information on the Web about popular entities.

| | Cumulative Resp. | | | | Url Ratings | | | |
|---|---|---|---|---|---|---|---|---|
| | Not Cov. | | Covered | | Not Cov. | | Covered | |
| **Domain** | Y | N | Y | N | Y | N | Y | N |
| country | 274 | 26 | 274 | 26 | 28 | 2 | 28 | 2 |
| country_travel | 238 | 8 | 258 | 8 | 25 | 0 | 27 | 0 |
| nba player | 269 | 53 | 223 | 43 | 33 | 0 | 27 | 0 |
| company | 280 | 48 | 228 | 40 | 31 | 2 | 26 | 1 |
| university | 309 | 31 | 235 | 25 | 34 | 0 | 26 | 0 |
| mountain | 242 | 38 | 282 | 38 | 26 | 2 | 30 | 2 |
| Total | 1612 | 204 | 1500 | 180 | 177 | 6 | 164 | 5 |

Table 4: User responses indicating whether the pages retrieved by aspects are relevant to the query.

#### 6.4.2 Relevant Information

To establish that the more information retrieved by aspects are in fact relevant to the user's information needs, we conducted an Amt user study. For each query $q$, we considered the top 10 aspects and for each aspect we consider the top retrieved web page, i.e., $D_1^a$. We constructed a list of aspect-based results $L_A$ which contained 4 results (selected at random from $D_1^a$), such that 2 overlapped with $D_{500}$ and 2 that did not overlap with $D_{500}$. Users were asked to evaluate if each of the results were (a) relevant, or (b) irrelevant to the original query $q$. In order to place the results in context, we also showed $L_A$ alongside $L_G$, the top 5 regular search engine results (these were not to be rated, but only for context).

We considered all 90 test queries with responses from 10 users in each case. The detailed results are shown in Table 4. The columns **Y** and **N** indicate whether the web pages were deemed relevant or not. The **Covered** and **Not Covered** columns separately consider the web pages in $L_A$ that were covered in $D_{500}$ and those that were not. The **Cumulative Responses** columns aggregate the responses for all users, while the **Url Ratings** columns aggregate the ratings of all users separately for each web page in $L_A$. As can be seen, in total there were 177 web pages that were not covered, but were deemed relevant by a majority of users.

The results indicate that overall, the vast majority of additional web pages retrieved by aspects were deemed relevant by a majority of users. In addition, the ratio of relevant to not-relevant results is about the same for the covered and not-covered web pages. This indicates that not only is the additional information relevant, but it is likely to be as relevant as the covered information.

Note that our coverage results also establish that our aspects are likely to span much more of an information space than alternate schemes that rely only on analyzing the results of the original query, e.g., the cluster labels in Clusty.

### 6.5 Comprehensive Performance Comparison

We compare the overall performance of Aspector to Grs, Ccl, and Bns by conducting a user study using Amt. We separately compared Aspector against each of the other systems. In each case, we selected a random subset of around 30 queries from our original set of 90 queries. We filtered the queries which don't return aspects from both systems. For each query, two lists of (at most) 8 aspects were generated, one using Aspector and the other using Grs, Ccl or Bns, which were then presented to an Amt rater with the following instructions:

*Each query represents the start of a session to explore more information on some topic (presumably related to the query). For each user query, we display two lists of other related queries and/or properties. We want you to compare the two lists for each query and identify which of the two lists enables a better subsequent exploration of information.*

The lists for each query were presented side-by-side, and the user could rate one to be better than the other, or simply rate them to be about the same. The raters were not informed of the source of each list and the side-by-side positioning of the lists was randomly selected. We collected responses from 15 raters for each query.

Tables 5, 6 and 7 summarize the results of the user study. The **Cumulative Responses** columns aggregate the responses for all raters for all queries. The **F**, **E**, and **A** columns indicate ratings in *favor* of Aspector, *even* ratings, and *against* Aspector (and in favor of Grs or Ccl) respectively. The **Query Ratings** columns aggregate the ratings of all the raters for each query, with an **F** indicating that more raters rated Aspector in favor of the other systems (respectively **E** and **A**).

As can be seen in Table 5, Aspector clearly outperforms Grs. The improvements are most likely due to its increased orthogonality and the grouping of aspects by vertical category. As is also clear in Table 6, Aspector also significantly outperforms Ccl, most likely due to its increased coverage.

To ascertain the statistical significance of the evaluation, for each comparison, we performed the standard *paired t-test.* For each individual query, we considered the total number of F responses against the A responses. For the comparison against Grs, the mean per-query difference (F-A) was 10.7, i.e., on average 10.7 out of the total 15 evaluators rated Aspector to be better. The difference was statistically significant with a two-tailed p-value less than 0.0001. For the comparison with Ccl, the mean difference was 13.1 and again significant with a p-value less than 0.0001.

Bns produces aspects for a small number of domains. To set the context for our comparison we measured the breadth of Bns. We chose the top 100 Wikipedia Infobox classes, and selected 15 entities (5 popular, 5 less common, and 5 randomly) from each class as in section 5.3. Bns provided aspects for 17.6% of the entities. In particular, Bns provided aspects for 29.4% of the popular entities and 9.8% for those less common entities. Bns provided no aspects for entities of 48 of the classes, including "scientist", "magazine" and "airport". The second limitation of Bns is that it only provides aspects for entity queries. Hence, Bns does not provide aspects for queries such as "vietnam travel", "seattle coffee" or "boston rentals". The third limitation of Bns is that it provides only class-level aspects, though the aspects may differ slightly from one instance to another. For example, Bns misses the aspect "starbucks" for "seattle", "turkey slap" for "turkey", and "number change" for "kobe bryant".

Of our 90 queries, only 41 of them obtain aspects from Bns. Table 7 shows that Aspector and Bns are rated comparably w.r.t. this limited set of queries. The advantages of Aspector come from the fact that it judiciously balances instance-level and class-level aspects. It is interesting to point out that when raters are not familiar with a particular instance, they tend to prefer the class-level aspects. In our experiment, this observation sometimes gives Bns an advantage.

| Domain | Cumulative Resp. | | | Query Ratings | | |
|---|---|---|---|---|---|---|
| | F | E | A | F | E | A |
| country | **60** | 7 | 8 | **4** | 0 | 1 |
| country_travel | **54** | 17 | 3 | **5** | 0 | 0 |
| nba player | **68** | 5 | 2 | **5** | 0 | 0 |
| company | **51** | 14 | 10 | **4** | 0 | 1 |
| university | **59** | 11 | 5 | **5** | 0 | 0 |
| mountain | **58** | 15 | 1 | **5** | 0 | 0 |
| Total | **350** | 69 | 29 | **28** | 0 | 2 |

Table 5: User responses comparing Aspector against Grs.

| Domain | Cumulative Resp. | | | Query Ratings | | |
|---|---|---|---|---|---|---|
| | F | E | A | F | E | A |
| country | **57** | 8 | 9 | **4** | 1 | 0 |
| country_travel | **62** | 5 | 0 | **5** | 0 | 0 |
| nba player | **69** | 3 | 2 | **5** | 0 | 0 |
| company | **56** | 1 | 12 | **5** | 0 | 0 |
| university | **62** | 3 | 8 | **5** | 0 | 0 |
| mountain | **53** | 7 | 7 | **5** | 0 | 0 |
| Total | **359** | 27 | 38 | **29** | 1 | 0 |

Table 6: User responses comparing Aspector against Ccl.

### 6.6 Instance-Level Versus Class-Level Aspects

Recall that Aspector balances between class-level and instance-level aspects for a given query. Consider, for example, the class of NBA players. There are 1365 players identified on Wikipedia. We were able to identify 8 or more candidate instance-level aspects for only 126 of them (9.2%). For 953 (69.8%) players, we were unable to infer any instance-level aspects. However there are 54 class-level aspects that appear with at least 5 instances, thus giving us a potentially larger pool of good candidate aspects. By balancing these two sources of aspects, Aspector is able to successfully compute reasonable aspects even for less frequent queries.

| Domain | Cumulative Resp. | | | Query Ratings | | |
|---|---|---|---|---|---|---|
| | F | E | A | F | E | A |
| country | **65** | 29 | 56 | **5** | 1 | 4 |
| nba player | 52 | 12 | **71** | 3 | 0 | **6** |
| company | 20 | 2 | **53** | 0 | 0 | **5** |
| university | **92** | 9 | 19 | **8** | 0 | 0 |
| mountain | 6 | 6 | **18** | 1 | 0 | 1 |
| Total | **235** | 58 | 217 | **17** | 1 | 16 |

Table 7: User responses comparing Aspector against Bns. Note this is the result after filtering 54% testing queries when Bns provides no aspects, in which case users always rate Aspector better.

| **Domain** | **Cumulative Resp.** | | | **Query Ratings** | | |
|---|---|---|---|---|---|---|
| | F | E | A | F | E | A |
| country | 12 | 17 | **46** | 0 | 0 | **5** |
| country_travel | 12 | 11 | **52** | 0 | 0 | **5** |
| nba player | 10 | 14 | **51** | 0 | 0 | **5** |
| company | 12 | 13 | **50** | 0 | 0 | **5** |
| university | 17 | 9 | **49** | 1 | 0 | **4** |
| mountain | 10 | 14 | **51** | 0 | 0 | **5** |
| Total | 73 | 78 | **299** | 1 | 0 | **29** |

Table 8: User responses when comparing Aspector with $K = 0$ and $K = 1$.

We compared the extent to which class-based aspect propagation contributes to the quality of aspects generated. For this, we again performed an Amt user study. We considered different values for the parameter $K$ in Formula 1: 0, 0.1, 1, 10, and 100, each indicating progressively higher contribution of class-level aspects. Aspect lists were generated for a subset of 30 queries (5 from each set) for each value of $K$. We compared two aspect lists at a time, and performed three sets of experiments comparing (1) $K = 0$ with $K = 1$ (Table 8), (2) $K = 0.1$ with $K = 10$ (Table 9), and $K = 1$ with $K = 100$ (Table 10). Each experiment used the same set of 30 queries and users were asked to pick which of the two sets of aspects they preferred for each query (same task description as in Section 5). Responses were collected from 15 users in each case. Note that to ensure significant differences in the aspect lists being compared, our experiments did not consider consecutive $K$ values (e.g., 0 and 0.1). In an earlier experiment where consecutive $K$ values were used, we found many queries have only a subtle difference and hence large numbers of users rated the lists to be comparable. The number of queries are fewer in Table 9 and Table 10, since the remaining ones resulted in the same aspect lists for both $K$ values – this is not surprising since, larger $K$ values result in the increased influence of the same set of class-based aspects.

We find that the aspect lists for $K = 1$ are rated significantly better than for $K = 0$ (the mean per-query difference (F-A) was 7.53 and significant with a two-sided p-value less than 1E-9). The lists for $K = 10$ were preferred to those with $K = 0.1$ (though by a smaller (F-A) of 2.4 with p value about 0.008), while the lists for $K = 100$ and $K = 1$ were rated about the same (the mean (F-A) was 0.3 and insignificant with p value about 0.8). Each of these seem to indicate clearly that class-based aspects are helpful in improving user experience.

However, we note that our results might marginally over-state the importance of class-based aspects. This is because a user's perception of aspects is dependent upon the user's interest and familiarity with the entity in question – if an entity, though popular, is not too familiar to a participant in the study, they are likely to select the class-based aspects. On the other hand, we found that for universally well known entities, such as the company microsoft, the lists with more instance-based aspects were always preferred.

| Domain | Cumulative Resp. | | | Query Ratings | | |
|---|---|---|---|---|---|---|
| | F | E | A | F | E | A |
| country | 11 | 15 | **34** | 0 | 0 | **4** |
| country_travel | 21 | **23** | 16 | **2** | 1 | 1 |
| nba player | 19 | 21 | **35** | 1 | 0 | **4** |
| company | 17 | 24 | **34** | 1 | 0 | **4** |
| university | 22 | 26 | **27** | 2 | 0 | **3** |
| mountain | 13 | 11 | **21** | 1 | 0 | **2** |
| Total | 103 | 120 | **167** | 7 | 1 | **18** |

Table 9: User responses when comparing ASPECTOR with $K = 0.1$ and $K = 10$.

| Domain | Cumulative Resp. | | | Query Ratings | | |
|---|---|---|---|---|---|---|
| | F | E | A | F | E | A |
| country | 3 | **8** | 4 | 0 | 0 | **1** |
| country_travel | 16 | **18** | 11 | **2** | 0 | 1 |
| nba player | 5 | 12 | **28** | 0 | 0 | **3** |
| company | **33** | 29 | 13 | **4** | 0 | 1 |
| university | 14 | **20** | 11 | 1 | 1 | 1 |
| mountain | 21 | 23 | **31** | 2 | 0 | **3** |
| Total | 92 | **110** | 98 | 9 | 1 | **10** |

Table 10: User responses when comparing ASPECTOR with $K = 1$ and $K = 100$.

### 6.7 Eliminating and Grouping Aspects

We now consider the impact of the content-based clustering that is used to identify duplicate aspects, and the vertical-category-based clustering that groups aspects belonging to the same category.

#### 6.7.1 Duplicate Elimination

When computing candidate aspects from query logs, it is possible to find multiple aspects that have different names, but are semantically the same. Such aspects have to be eliminated in order for the summary to cover more distinct axes. As explained in Section 5.1, ASPECTOR applies a graph partitioning algorithm that only has a single parameter, the similarity threshold $\sigma$. We conjecture that this similarity threshold is more intuitive to set and is stable across different domains.

To test our hypothesis, we randomly selected 5 queries each from 5 domains. For each query, we took the top 30 aspects candidates and manually created a gold-standard with the correct aspect clustering results. We computed aspect lists for the queries with different values for the threshold $\sigma$ and compared the results against the gold-standard.

We use the *F-Measure* ($F$) to evaluate the clustering results. In particular, we view clustering as a series of decisions, one for each of the $N(N-1)/2$ pairs of aspects.

Figure 4 plots the $F$ values for different values of the threshold $\sigma$ for the five test domains. We found that in each case the best performance is between threshold values 0.25 and 0.4. The results indicate that clustering performance with respect to $\sigma$ is pretty stable across domains. Hence, in all our experiments, we set a single value $\sigma = 0.35$.

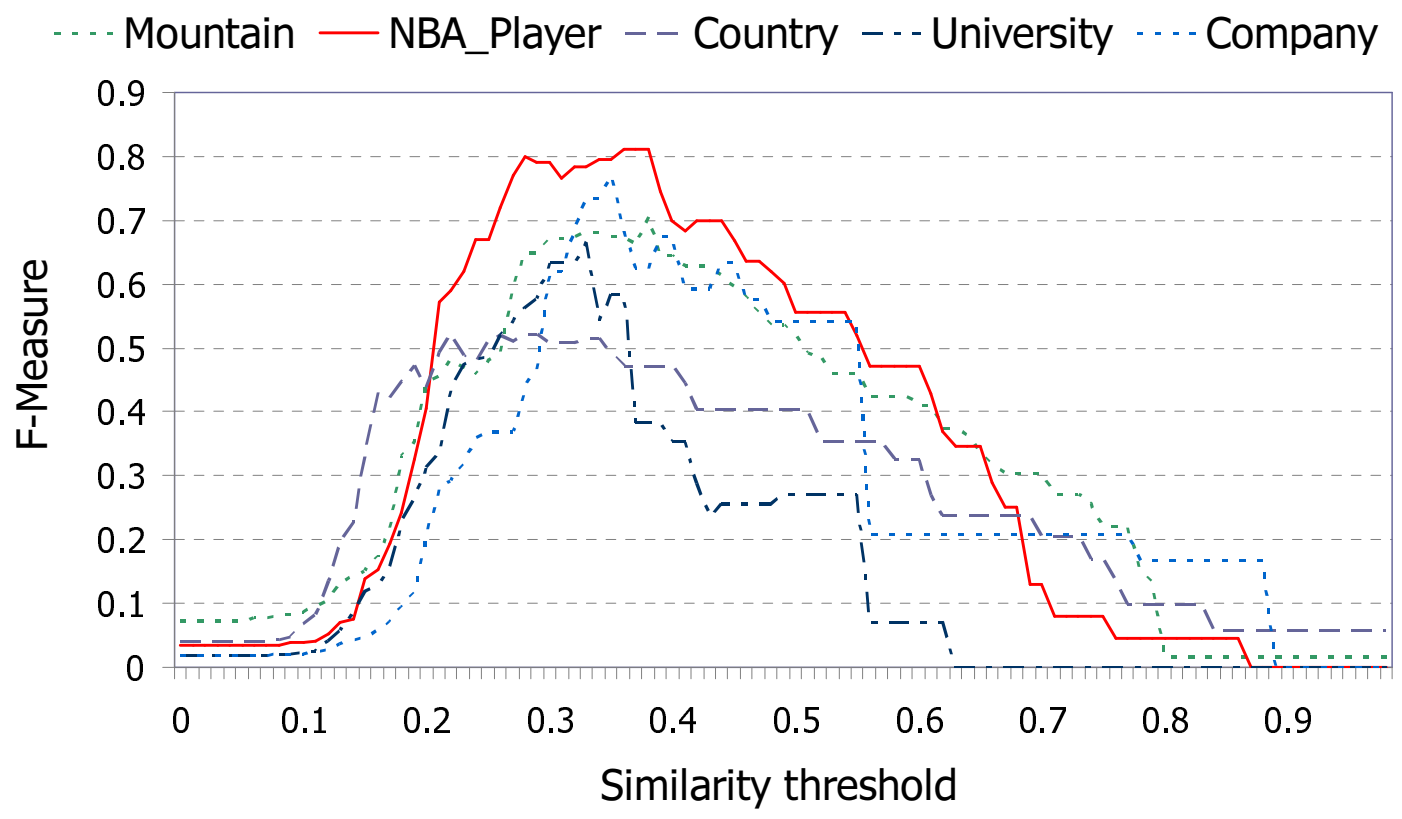

Figure 4: F-Measure of aspect clustering for different values of the similarity threshold $\sigma$. In each domain, the best performance is around 0.35.

| Domain | Cumulative Resp. | | | Query Ratings | | |
|---|---|---|---|---|---|---|
| | F | E | A | F | E | A |
| company | **25** | 14 | 6 | **3** | 0 | 0 |
| university | **18** | 7 | 5 | **2** | 0 | 0 |
| mountain | 6 | 3 | 6 | 0 | **1** | 0 |
| Total | **49** | 24 | 17 | **5** | 1 | 0 |

Table 11: User responses comparing Aspector with vertical-category grouping and without. $F$, $E$, and $A$ are responses in favor, even, and against grouping.

### 6.7.2 Vertical-Category Based Grouping

In addition to duplicates, we observed that often multiple aspects might belong to the same vertical category. Rather than represent each as a separate aspect, we can summarize the aspects by presenting them as a single group. Note that grouping does not *eliminate* the aspects, but simply lists all of them as a single aspect. For 6 out of the 90 queries in our dataset, Aspector was able to group aspects by vertical category.

As before, we deployed an Amt user study for the 6 test queries and the results are shown in Table 11, with **F** indicating the number of responses in favor of vertical grouping (with **E** and **A** defined accordingly). As can be seen, the aspect lists with grouping were favored in comparison to the ones without grouping.

Currently, Aspector groups aspects of the same vertical category only if there is no corresponding disambiguation pages in Wikipedia. This conservative solution avoids potential errors when ambiguity exists, but also misses opportunities. For example, for the query mount bachelor, mount hood and mount baker appear as separate aspects since there are disambiguation pages for both entities. Refining the grouping condition is a rich topic for future work.

## 7. Related Work

We discuss work related to ours in the area of search-result organization and of query-log mining.

### 7.1 Search Result Organization

Several works have considered how to better organize search results. Agrawal, Gollapudi, Halverson, and Ieong (2009) classify queries and documents into categories and return search results by considering both document relevance and diversity of results. In contrast, Aspector computes more fine grained aspects instead of abstract categories for exploratory queries which are not necessary to be ambiguous. Some commercial systems like Kosmix and Yahoo!Glue categorize information based on type or format (e.g. photo, video, news and map) and retrieve top results for each category. Though the different types often approximate aspects, they do not represent a rich set of semantically different groups of information and are not sensitive to instance-specific aspects. The Carrot2 search engine applies text clustering techniques over returned search pages and extracts keywords to summarize each cluster. Similar works were done by Bekkerman, Zilberstein, and Allan (2007), Blei et al. (2003), Crabtree, Andreae, and Gao (2006), and Wang, Blei, and Heckerman (2008). Multi-faceted search (Yee, Swearingen, Li, & Hearst, 2003) organizes collections based on a set of category hierarchies each of which corresponds to a different facet. However the category hierarchies requires heavy human effort for construction and maintenance. The Correlator system from Yahoo! performs semantic tagging of documents to enable mining of related entities for a query. These algorithms don't necessary discover clusters which correspond to Web users' search interests, and it is difficult to generate informative cluster labels from documents. Our use of query logs complements such document-based approaches, but reflects searchers intentions rather the intentions of the publishers.

Wang and Zhai (2007) proposed to organize search results based on query logs. They represent each query as a pseudo-document enriched with clickthrough information and pick the top-k that are similar to the current query, and cluster them into aspects. Then, they classify each resulting page into corresponding aspect by similarity. In contrast, we generate aspects based on the idea of query refinements which don't require the aspect and current query to have similar clickthrough. For example, the query vietnam travel visa is an important aspect for vietnam travel, but won't have the same click-through properties.

### 7.2 Query-Log Mining

There have been several efforts to mine query logs for interesting artifacts. Pasca and Durme (2007) extract relevant attributes for classes of entities from query logs rather than from Web documents as done by Bellare et al. (2006). The main goal of these works is to create a knowledge base of entities, and hence their results are most appropriately compared with Wikipedia or Freebase.

Query refinement and suggestion analyze query logs to predict the next most probable query following the current query (Cucerzan & White, 2007; Jones, Rey, Madani, & Greiner, 2006; Kraft & Zien, 2004; Velez, Wiess, Sheldon, & Gifford, 1997). Hence, their goal is to help users find a single result page rather than help navigating a body of relevant information. Bonchi, Castillo, Donato, and Gionis (2008) proposed to decompose a query into a small set of queries whose union corresponds approximately to that of the original query. However, as our experiments illustrated, the constraint that the union of the resulting pages correspond approximately to that of the original query significantly limits the available body of information we expose to the user.

Wang et al. (2009) mine a set of global latent query aspects, and dynamically select top k aspects for a given query $q$ to help better navigate the information space. While in some ways this is similar to Aspector, there are two key differences. First, they discover the set of global latent query aspect via maximizing a target function, where the aspect set aims to apply to many classes of (important) queries. In contrast, Aspector applies class-based label propagation to identify the aspects. Therefore, the aspects tend to be more fine-grained and more query(class)-specific. Second, when selecting the k aspects for a query $q$, Wang et al. apply another optimization function which tries to cover as many original (frequent) query refinements of $q$. This works fine for popular queries but not for less popular queries which have few query refinements. Our experiments show that for most classes, there is a long tail of such less popular queries.

## 8. Conclusions

We described the Aspector system for computing aspects of web-search queries. Aspects are intended to offer axes along which the space of information relevant to a query can be organized, and therefore enable search engines to assist the user in exploring the space.

Aspector generates candidate aspects from query logs and balances aspects that are common to classes of entities vs. those that are specific to particular instances. Aspector also eliminates duplicate aspects and groups related aspects using a reference ontology. In contrast with a purely knowledge-based approach, Aspector's results are much broader and include aspects of interest to specific instances. In contrast with an approach based solely on clustering the results of the query, Aspector can include aspects that are not represented directly in the query's answer.

We set the weights of all edges from instances to classes uniformly when computing class-based aspects. A future direction is to compute better informed weighting functions based on available temporal, spatial and contextual constraints. Another future work is to allow multi-class memberships based on other ontologies besides Wikipedia Infobox.

To incorporate aspects into a mainstream search engine we need to address two challenges. First, we need to reliably identify from the query stream which queries benefit from a summarization approach. Some works (Miwa & Kando, 2007; White & Roth, 2009) have been conducted in this area, but much more needs to be investigated. Second, as done in Kosmix, we need to dynamically generate effective visualizations for aspects.